\documentclass[conference]{IEEEtran}

\ifCLASSINFOpdf
\else
\fi
\IEEEoverridecommandlockouts
\usepackage{stfloats,color}
\usepackage{amsfonts}
\usepackage[cmex10]{amsmath}
\usepackage{graphicx}
\usepackage{setspace}
\usepackage{cite}
\usepackage{bm}
\usepackage{array}
\usepackage{algorithmic}
\usepackage{mdwmath}
\usepackage{mdwtab}
\usepackage{amssymb}
\usepackage{caption}
\usepackage{subcaption}
\usepackage[font=footnotesize]{subfig}
\usepackage{fixltx2e}
\usepackage{url}
\usepackage{times}
\usepackage{epsfig}
\usepackage{latexsym}
\usepackage{epstopdf}
\usepackage{verbatim}
\usepackage{units}
\usepackage{amsthm}
\usepackage{mdwlist}
\usepackage{lipsum,amsmath,multicol}
\usepackage{soul}
\graphicspath{{./figures/}}
\newtheorem{thm}{Theorem}
\newtheorem*{prof}{Proof}
\newtheorem{lemm}{Lemma}

\begin{document}
%
\title{Achievable Degrees of Freedom Region of MIMO Relay Networks using Detour Schemes}

\author{\IEEEauthorblockN{ Ahmed A. Zewail, Mohammed Nafie}
\IEEEauthorblockA{Wireless Intelligent Networks Center (WINC)\\ Nile University, Giza, Egypt \\  
Email: ahmed.zewail@nileu.edu.eg, \\
\ mnafie@nileuniversity.edu.eg}
\and
\IEEEauthorblockN{Yahya Mohasseb}
\IEEEauthorblockA{Department of Communications \\ The Military Technical College \\ Cairo, Egypt 11331\\
Email: mohasseb@ieee.org}
\and
\IEEEauthorblockN{Hesham El Gamal}
\IEEEauthorblockA{ECE Department\\ The Ohio State University\\ Columbus, OH \\
Email: helgamal@ece.osu.edu}}

\maketitle

\begin{abstract}
In this paper, we study the degrees of freedom (DoF) of the MIMO relay networks. We start with a general Y channel, where each user has $M_i$ antennas and aims to exchange messages with the other two users via a relay equipped with $N$ antennas. Then, we extend our work to a general 4-user MIMO relay network. Unlike most previous work which focused on the total DoF of the network, our aim here is to characterize the achievable DoF region as well. We develop an outer bound on the DoF region based on the notion of one sided genie. Then, we define a new achievable region using the Signal Space Alignment (SSA) and the Detour Schemes. Our achievable scheme achieves the upper bound for certain conditions relating $M_i$'s and $N$.
\end{abstract}
\footnotetext[1]{This paper was made possible by NPRP grant $\#$ 4-1119-2-427 from the Qatar National Research Fund (a member of Qatar Foundation). The statement made herein are solely responsibility of the authors.}
\footnotetext[2]{Ahmed A. Zewail is now with the Department of Electrical
Engineering, Pennsylvania State University, University Park, PA 16802 USA.}
\footnotetext[3]{Mohammed Nafie is also affiliated with the Department of Electronics and Communications, Cairo University.}



%
\IEEEpeerreviewmaketitle
\vspace{-.1 in}
\section{Introduction}
\IEEEPARstart{C}\small{o}operative communications for wireless networks have gained great research interest, due to its ability to enhance the performance of wireless networks. The relay networks, specially the MIMO relay networks, where an additional node acting as a relay is supporting the exchange of information between the network users, has attracted an extensive research attention. Since characterizing the capacity of the MIMO relay networks is very complex, a lot of work was done to study the relay networks using an alternative metric, which is the degrees of freedom (DoF).\newline
The DoF of the MIMO Y channel was investigated in \cite{lee2010degrees,she2012degrees,chaaban2013degrees}. The Signal Space Alignment (SSA) was presented in \cite{lee2010degrees} to achieve total DoF $= 3M$, for $N\geq \lceil 1.5M \rceil$, where $M$ and $N$ are the number of antennas at each user and the relay, respectively. The SSA is a network coding technique, which is similar to the interference alignment (IA) \cite{jafar2008degrees} from one side. That is, each of them makes an efficient use of dimension of the signal space. However, the IA attempts to overlap all interference signals to minimize the dimension of the signal space they occupy, while the SSA focuses on choosing the beamforming matrices, such that "bidirectional" signal vectors, which correspond to the exchange between the two users but are sent via the relay, are aligned at the relay in the same subspace. \newline
In \cite{she2012degrees}, the authors targeted a generalization of the work done in \cite{lee2010degrees}, by considering a Y channel, where each user has $M_i$ antennas, while the relay has $N$ antennas. Using the SSA, they proved the achievability of total DoF $= m_1+m_2+m_3$, where $m_i$ is the total number of signals transmitted from user $i$, under the conditions $N\geq 0.5(m_1+m_2+m_3)$ and $M_i+M_j\geq N+n_{ij}$, where $n_{ij}$ is the number of signals between nodes $i$ and $j$. Then, the authors in \cite{chaaban2013degrees}, completely characterized the total DoF of the MIMO Y channel, with no constraints on $N$. Again, using the SSA, they achieved a total DoF $= \min(2N,2M_2+2M_3,M_1+M_2+M_3)$, where $M_1\geq M_2 \geq M_3$. They derived the outer bound of the DoF based on the notion of one sided genie.\newline
The total DoF for a 4-user relay network was studied in \cite{tiansignal}. The 4 users were split into two clusters, such that the two users within each cluster only communicate with each other. 
 Each user has $M$ antennas, while the relay has $N$ antennas. Using a combination of SSA and TDM, it was shown that a total DoF $= 2\min(2M,N)$ is achievable. Then, this work was extended to the $L$-cluster, $K$-user MIMO multi-way relay channel with no direct links, where message exchanges were limited to users within the same cluster \cite{tian2013degrees}. \\\\   
The capacity of different topologies of the deterministic relay networks with no direct links was studied in \cite{mokhtar2010deterministic,zewail2013deterministicitw,zewail2013deterministic}. The authors developed a new upper bound on the capacity region based on the notion of the single sided genie, then they proved the achievability of this upper bound by using the \textit{Detour Schemes (DS)}, where some bits are sent via alternative paths instead of sending them directly. Note that a similar outer bound was developed in \cite{tian2013degrees} for MIMO relay networks; which will be used here.\\ 

  In our work, we study the DoF of the MIMO relay networks. We focus on showing the role of the Detour Schemes, which we used before to achieve the capacity of the deterministic relay networks, in achieving the DoF region of the MIMO relay networks \cite{mokhtar2010deterministic,zewail2013deterministic,zewail2013deterministicitw}.
 First, we study the MIMO Y channel. We develop a new outer bound on the DoF region based on the notion of one sided genie. Then, we define a new achievable DoF region using the Signal Space Alignment and the Detour Scheme. 
Subsequently, we extend our results to the 4-user MIMO relay networks. 
This work can be considered a generalization of most of the above work, as we consider a general 4-user relay network not divided into clusters as in \cite{tiansignal}, and we study the achievability of the DoF region, not only the total DoF as in \cite{lee2010degrees,she2012degrees,chaaban2013degrees}, \cite{tiansignal} and \cite{tian2013degrees}.\newline \\
The rest of this paper is organized as follows: In Section \ref{system_model}, we describe our network and the main assumptions. Then, we study the DoF region of the MIMO Y channel in Section \ref{y_channel}, where we define an outer bound on the DoF region based on the notion of single sided genie and a new achievable DoF region via the Signal Space Alignment and the Detour Scheme. Subsequently, we extend our results to the 4-user MIMO relay networks in Section \ref{4_channel}. Then, The development of the outer bound based on the notion of single sided genie developed in \cite{tian2013degrees} and \cite{mokhtar2010deterministic} is briefly explained in Section \ref{upperboundsection}. Numerical examples that illustrate our achievable schemes are presented in Section \ref{ex}. Finally, our conclusions are stated in Section \ref{concl}.  
\section{System Model}\label{system_model}
\begin{figure}
\includegraphics[width=0.45\textwidth,height=0.15\textheight]{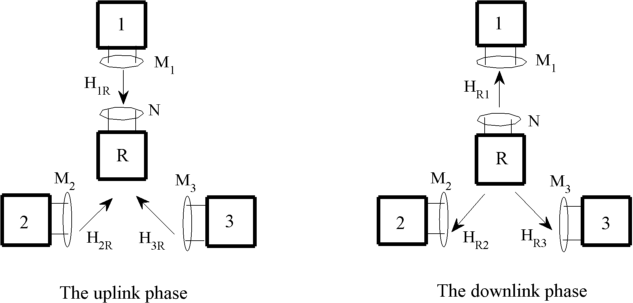}
\centering
\caption{\small The MIMO 3-user relay network}\label{fig:sm3}
\end{figure}

\begin{figure}
\includegraphics[width=0.45\textwidth,height=0.15\textheight]{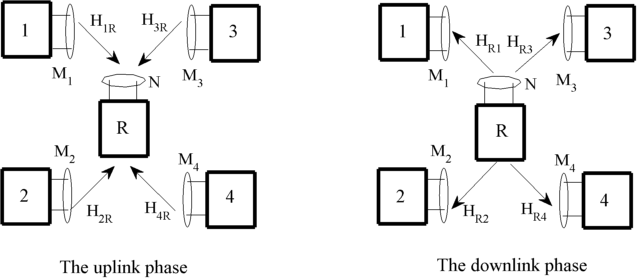}
\centering
\caption{\small The MIMO 4-user relay network}\label{fig:sm4}
\end{figure}

We consider a $K$-user MIMO relay network with no direct links. We study two cases: $K=3$, which corresponds to the Y channel shown in Fig. \ref{fig:sm3} and $K=4$, which corresponds to a 4-user relay network, as shown in Fig. \ref{fig:sm4}. User $i$ is equipped with $M_i$ antennas, and can exchange private messages with other network users via a relay equipped with $N$ antennas. We define the number of used antennas at the relay and user $i$ as $\bar N$ and $\bar M_i$ respectively. For both networks, we assume without loss of generality that the nodes are labeled in order of descending number of antennas. This implies that $M_1 \geq M_2 \geq M_3 (\geq M_4)$. \\
We consider a full duplex scenario, where the transmission takes place over two phases: uplink and downlink. In the uplink phase, each user transmits its messages to the relay, therefore the received signal at the relay is given by   
\begin{equation}\nonumber
\bm{y_r(t)} = \sum_{i=1}^{K} \mathbf{H_{iR}}\bm{x_i(t)}+\bm{z_r(t)}
\end{equation}
where $\bm{y_r(t)}$ is a $\bar N \times 1$ vector, $\mathbf{H_{iR}}$ is a $\bar N \times \bar M_i$ random channel matrix from user $i$ to the relay, $\bm{x_i(t)}$ is a $\bar M_i \times 1$ vector representing the transmitted signal from node $i$, $t$ is the time index, and $\bm{z_r(t)}$ is an i.i.d. Gaussian noise vector, i.e. ($\bm{z_r}\backsim\mathcal{N}(0,I)$). \\
In the downlink phase, the received signal at user $i$ from the relay is given by: 
\begin{equation}\nonumber
\bm{y_i(t)} = \mathbf{H_{Ri}}\bm{x_r(t)}+\bm{z_i(t)}
\end{equation}
where $\bm{y_i(t)}$ is a $\bar M_i \times 1$ vector, $\mathbf{H_{Ri}}$ is a $\bar M_i \times \bar N$ random channel matrix from the relay to user $i$, $\bm{x_r(t)}$ is a $\bar N \times 1$ vector representing the transmitted signal from the relay, and $\bm{z_i(t)}$ is an i.i.d. Gaussian noise vector, i.e. ($\bm{z_i}\backsim\mathcal{N}(0,I)$). \\

\section{The MIMO Y Channel}\label{y_channel}
In this section, we study the DoF region of a general MIMO Y channel. We expand the range of achievable DoF beyond that found in \cite{chaaban2013degrees}. Therefore, our work here can be considered as an extension of the work in \cite{chaaban2013degrees}, where the authors only focused on the total DoF and proved its achievability. 
\begin{thm}
An outer bound on the degrees of freedom (DoF) region of the MIMO Y channel is given by the following inequalities:
\begin{equation}\label{yUL1}
d_{ij}+d_{ik}\leq \min(N,M_i,M_j+M_k)
\end{equation}
\begin{equation}\label{yDL1}
d_{ji}+d_{ki}\leq \min(N,M_i,M_j+M_k)
\end{equation}
\begin{equation}\label{yUL2}
d_{ij}+d_{ik}+\max(d_{jk},d_{kj})\leq \min(N,M_j+M_k)
\end{equation}
\begin{equation}\label{yDL2}
d_{ji}+d_{ki}+\max(d_{jk},d_{kj})\leq \min(N,M_j+M_k)
\end{equation}
where $d_{ij}$ is the degrees of freedom between user $i$ to user $j$, and $\{i,j,k\} \in \{1,2,3\}$.\\
Also, the total degrees of freedom $D_t$ is bounded by
\small 
\begin{equation}\label{yDt}
D_t = {\sum_{i=1}^{3}\sum_{j=1, i\neq j}^{3}} d_{ij}  \leq \min(2N,M_1+M_2+M_3,2M_2+2M_3)
\end{equation}
\normalsize
\end{thm}
The development of these inequalities will be explained briefly in Section \ref{upperboundsection}. \newline 
Now, we will derive a new achievable region using a combination of two schemes: The Signal Space Alignment (SSA) \cite{lee2010degrees} and the Detour Scheme (DS), which was used to achieve the capacity region of the 3-user relay network stated in Theorem 2 in  \cite{zewail2013deterministicitw}.\\
Let $d_{ij}^*=\max(d_{ij},d_{ji})$, then the number of used antenna at user $i$ and the relay, respectively is given by: $\bar M_i = \max(d_{ij}+d_{ik},d_{ji}+d_{ki})$, and $\bar N =d_{12}^*+d_{13}^*+d_{23}^*$. From (\ref{yUL1}) and (\ref{yDL1}), it is clear that $\bar M_i \leq M_i$, however, $\bar N$ may be greater than $N$.
\subsection{Signal Space Alignment ($\bar N\leq$ $N$)}\label{ySSA}
The SSA is sufficient to achieve all DoF tuples that satisfy the conditions in Theorem 1, with $\bar N \leq N$. The key idea behind the SSA is that we design the beamforming matrices at the transmitting users such that the "bidirectional" signal vectors, correspond to the exchange between two users but both are sent via relay, are aligned at the relay to the same subspace. Thus, the transmitted signal at user $i$ is constructed as 
\begin{equation}\nonumber
\bm{x_i}=\sum_{j=1, j\neq i}^{K} \mathbf{V_{ij}}\bm{s_{ij}}
\end{equation} 
where $\bm{s_{ij}}$ is a $d_{ij} \times 1$ data vector sent from user $i$ to user $j$, $\mathbf{V_{ij}}$ is $\bar M_i \times d_{ij}$ beamforming matrix at user $i$ and $K=3$.\\
The relay receives a superposition of the transmitted signals from all network users, thus the received signal at the relay is given by
\begin{equation}\label{receivedyrelay}\nonumber
\bm{y_r}= \sum_{i=1}^{K}\mathbf{H_{iR}}\bm{x_i}+\bm{z_r}
\end{equation}  
where $\mathbf{H_{iR}}$ is  $\bar N \times \bar M_i$ matrix which represents the channel between user $i$ and the relay. \\ 
According to the SSA, we need to design the beamforming matrices $\mathbf{V_{ij}}$ and $\mathbf{V_{ji}}$ such that users $i$ and $j$ exchange their messages in the same subspace. If we assume $d_{ij}\geq d_{ji}$, then we should have the following condition
\begin{equation}\nonumber
span(\mathbf{{H_{jR}}V_{ji}}) \subseteq span(\mathbf{{H_{iR}}V_{ij}})
\end{equation} 
and $\mathbf{{H_{jR}}V_{ji}}$ has full column rank.\\
At the relay, we use zero forcing to separate different pairs of messages. To extract the messages between users $i$ and $j$, the relay multiplies the received signal by the matrix $\mathbf{D_{ij}}$ with size $d_{ij}^* \times \bar N$. The rows of this matrix span the null space of the channel between the remaining user and the relay, $\mathbf{H_{kR}}$, hence:
\begin{equation}\nonumber
\mathbf{D_{ij}H_{kR}}=\bm{0}
\end{equation}
Thus, the relay obtains the $d_{ij}^*$ vector $\bm{w_{ij}}$ which represents a combination of the messages between users $i$ and $j$:
\begin{equation}\nonumber
\bm{w_{ij}}=\mathbf{D_{ij}H_{iR}V_{ij}}\bm{s_{ij}}+\mathbf{D_{ij}H_{jR}V_{ji}}\bm{s_{ji}}+\mathbf{D_{ij}}\bm{z_r}
\end{equation}
For each pair of users $(i,j)$ the relay calculates the corresponding zero forcing matrix $\mathbf{T_{ij}}$, and forms the transmitted signal vector $\bm{x_r}$ given by:
\begin{equation}\label{transmitedyrelay}
\bm{x_r}={\sum_{i=1}^{K}\sum_{j>i}^{K}}\mathbf{T_{ij}}\bm{w_{ij}}
\end{equation}
where $\mathbf{T_{ij}}$ is $\bar N \times d_{ij}^*$ zero-forcing matrix, such that $\mathbf{{H_{Rk}} T_{ij}}= \bm{0}$. Now, the received signal at user $i$ is given by 
\begin{equation}\nonumber
\bm{y_i} = \mathbf{H_{Ri}} \begin{bmatrix}
      \mathbf{T_{ij}} & \mathbf{T_{ik}}
           \end{bmatrix}   \begin{bmatrix}
        \bm{w_{ij}} \\
      \bm{w_{ik}} 
           \end{bmatrix} +\bm{z_i}
\end{equation}
Since $\begin{bmatrix}
      \mathbf{T_{ij}} & \mathbf{T_{ik}}
           \end{bmatrix} $ is of full column rank, and $d_{ij}^*+d_{ik}^* \leq \bar N$,  user $i$ can decode $\bm{s_{ji}}$ and $\bm{s_{ki}}$ achieving $d_{ji}+d_{ki}$ DoF.  \\
Consequently, we can state the following Lemma:           
\begin{lemm}
The Signal Space Alignment can be used to achieve any integral DoF tuple that satisfies the conditions in Theorem 1, with $\bar N \leq N$.  
\end{lemm}
\subsection{Detour Scheme ($\bar N>N$ and $M_1\neq\min$($N$,$M_1$,$M_2+M_3$))}
Here we show that, for any DoF tuple that satisfies the conditions in Theorem 1, with $\bar N>N$ and $M_1\neq\min$($N,M_1,M_2+M_3$), we can use the Detour Scheme (DS) to convert our network into an equivalent one to which we can apply the SSA. \newline 
Depending on $\arg{\max(d_{ij} , d_{ji})}$ for $i,j$ $\in$ $\{1,2,3\}$, $\bar N$ can take eight different forms. However, since we assume $\bar N$= $d_{12}^*+d_{13}^*+d_{23}^*>N$ we have to exclude all forms that can result from (\ref{yUL2}) and (\ref{yDL2}) since they would necessarily imply $\bar N \leq N$. Therefore, $\bar N$ may only be in one of the following two forms:
\begin{equation}\label{ydetour1}
\bar N= d_{12}+d_{23}+d_{31}
\end{equation}  
\begin{equation}
\bar N= d_{21}+d_{13}+d_{32}
\end{equation}
It is clear that $\bar N$ represents the cycle between the network users in one of its two possible directions. In this case, the Detour Scheme can be used to route certain messages to their destinations via alternate routes involving other users, such that the SSA can be used to achieve the total DoF. To explain our DS, we assume that
\begin{equation}\label{yviolated}
d_{12}+d_{23}+d_{31} > N
\end{equation}
To apply the SSA, we need to force $\bar N$ to be less than or equal to $N$, thus we will subtract $\lambda$ from the LHS of (\ref{yviolated}), such that 
\begin{equation}
d_{12}+d_{23}+d_{31}-\lambda \leq N
\end{equation}
Now, the omitted $\lambda$-messages should be transmitted via alternative path (detour) to their respective destination. For example, if we detour $\lambda$-messages from $d_{23}$ via user 1, then each of $d_{21}$ and $d_{13}$ should be increased by $\lambda$. Thus, the DoF over the reverse cycle should be increased by $2\lambda$ as follows
\begin{equation}\nonumber
d_{21}+d_{13}+d_{32}\rightarrow d_{21}+d_{13}+d_{32}+2\lambda
\end{equation} 
It remains to show that the resulting network after modification can still achieve DoF using SSA. Once this is proven, we can state the following lemma:
\begin{lemm}
If $M_1\neq\min$($N,M_1,M_2+M_3$), it is possible to achieve any DoF tuple that satisfies the conditions in Theorem 1, with $\bar N>N$, by using the Detour Scheme to convert our network to an equivalent one that achieves the original DoF tuple via alternative paths.  
\end{lemm} 
\begin{prof}
See appendix A.
\end{prof}
\subsection{$\bar N>N$ and $M_1=\min$($N,M_1,M_2+M_3$)} In this case, the Detour Scheme fails to achieve this DoF tuple. We will verify this by the following example.\newline 
Consider a network with $(M_1,M_2,M_3,N)$ = (3,2,2,4) and $d$ $= (d_{12},d_{13},d_{21},d_{23},d_{31},d_{32})$ = (2,0,0,2,2,0). Since $\bar N=6=N+2$, we need to detour 2 messages, therefore we will subtract 2 from $d_{12}+d_{23}+d_{31}$ and add 4 to $d_{21}+d_{13}+d_{32}$, thus $D_t=8$. However, from (\ref{yDt}), we get $D_t \leq 7$. 
Therefore, it is unclear whether this DoF tuple can be achieved using other methods, or is not achievable.  
\section{The MIMO 4-user Relay Networks}\label{4_channel}
In this section, we extend our results to the MIMO 4-user relay networks. 
\begin{thm}
An outer bound on the degrees of freedom region of the 4-user relay network is given by the following inequalities:
\begin{equation}\label{onenode}
d_{ij}+d_{ik}+d_{il}\leq \min(M_i,N,M_j+M_k+M_l)
\end{equation}
\begin{equation}\label{onenode2}
d_{ji}+d_{ki}+d_{li}\leq \min(M_i,N,M_j+M_k+M_l)
\end{equation}
\begin{equation}\label{twoUL}
d_{ik}+d_{il}+d_{jk}+d_{jl}+\max(d_{ij},d_{ji})\leq \min(N,M_i+M_j)
\end{equation}
\begin{equation}\label{twoDL}
d_{ki}+d_{li}+d_{kj}+d_{lj}+\max(d_{ij},d_{ji})\leq \min(N,M_i+M_j)
\end{equation}
\begin{multline}\label{threeUL}
d_{il}+d_{jl}+d_{kl}+d_{ij}+d_{ik}+\max(d_{jk},d_{kj})\\  \leq \min(N,M_i+M_j+M_k)
\end{multline}
\begin{multline}\label{threeDL}
d_{li}+d_{lj}+d_{lk}+d_{ji}+d_{ki}+\max(d_{jk},d_{kj})\\ \leq \min(N,M_i+M_j+M_k)
\end{multline}
where $d_{ij}$ is the degrees of freedom between user $i$ to user $j$, and $\{i,j,k,l\} \in \{1,2,3,4\}$.\\
Also, the total degrees of freedom is bounded by
\small
\begin{multline}\label{Dt}
D_t = {\sum_{i=1}^{4}\sum_{j=1, i\neq j}^{4}} d_{ij} \\ \leq \min(2N,M_1+M_2+M_3+M_4,2M_2+2M_3+2M_4)
\end{multline}
\normalsize
\end{thm}  
Now, we will define a new achievable region using a combination of two schemes: The Signal Space Alignment (SSA) \cite{lee2010degrees} and two Detour Schemes (DS), which were used to achieve the capacity region of the 4-user relay network stated in Theorem 1 in \cite{zewail2013deterministic}.\\
Again, let $d_{ij}^*=\max(d_{ij},d_{ji})$, $\bar M_i = \max(d_{ij}+d_{ik}+d_{il},d_{ji}+d_{ki}+d_{li})$, and $\bar N =d_{12}^*+d_{13}^*+d_{14}^*+d_{23}^*+d_{24}^*+d_{34}^*$. From (\ref{onenode}) and (\ref{onenode2}), it follows that $\bar M_i \leq M_i$. However, $\bar N$ may be greater than $N$.

\subsection{The Signal Space Alignment ($\bar N\leq$ $N$)}
The SSA is sufficient to achieve all DoF tuples that satisfy the conditions in Theorem 2, with $\bar N \leq N$. The scheme will follow the same procedure explained in subsection \ref{ySSA} but for $K=4$. Therefore, the signal transmitted by the relay, $\bm{x_r}$, will be given by equation (\ref{transmitedyrelay}) with K=4.\\ 
Finally, the received signal at user $i$ is given by 
\begin{equation}\nonumber
\bm{y_i} = \mathbf{H_{Ri}} \begin{bmatrix}
      \mathbf{T_{ij}} & \mathbf{T_{ik}} &  \mathbf{T_{il}}
           \end{bmatrix}  \begin{bmatrix}
        \bm{w_{ij}} \\
      \bm{w_{ik}} \\
      \bm{w_{il}} 
           \end{bmatrix} +\bm{z_i}
\end{equation}
Since $\begin{bmatrix}
      \mathbf{T_{ij}} & \mathbf{T_{ik}} &  \mathbf{T_{il}}
           \end{bmatrix} $ is of full column rank, because $d_{ij}^*+d_{ik}^*+d_{il}^*\leq \bar N$, then user $i$ can decode $\bm{s_{ji}, \bm{s_{ki}}}$ and $\bm{s_{li}}$ achieving $d_{ji}+d_{ki}+d_{li}$ DoF.  \\
Consequently, we can state the following Lemma: 
\begin{lemm}
The Signal Space Alignment can be used to achieve any integral DoF tuple that satisfies the conditions in Theorem 2, with $\bar N \leq N$.  
\end{lemm}
\hspace{-.24 in}
\textbf{\textit{Remark.}} We can prove the achievability of the total DoF for different cases using the SSA, by choosing $d_{ij}=d_{ji}$ such that $D_t=\min(2N,M_1+M_2+M_3+M_4,2M_2+2M_3+2M_4)$.
\subsection{The Detour Schemes}
If we consider a case, where a DoF tuple satisfies the conditions in Theorem 2, with $\bar N > N$, then $\bar N$ cannot be in the form of the LHS's of (\ref{threeUL}) nor (\ref{threeDL}). Therefore, $\bar{N}$ will be in one of the two following formulas: 
\begin{equation}\label{eqndet31}
\bar{N}=d_{ij}+d_{jk}+d_{ki}+\max(d_{li}+d_{lj}+d_{lk},d_{il}+d_{jl}+d_{kl})
\end{equation}
\begin{equation}\label{eqndet41}
\bar N = d_{ij}+d_{jk}+d_{kl}+d_{li}+\max(d_{jl},d_{lj})+\max(d_{ik},d_{ki})
\end{equation} 
where $\{i,j,k,l\} \in \{1,2,3,4\}$.\\

We can observe that the set of combinations represented by (\ref{eqndet31}) contains a 3-node cycle corresponding to the data flow in the first three terms, while the combinations represented by (\ref{eqndet41}) will contain two 3-node cycles. These two cycles will depend on the two maximizations in its RHS. For example, if $\max(d_{jl},d_{lj})+\max (d_{ik},d_{ki})$ = $d_{lj}+d_{ik}$, then the 3-node cycles are $i, k, l$ obtained from the terms  $d_{ik}, d_{kl}$ and $d_{li}$, and the cycle  $l, j, k$ obtained from the terms $d_{lj}, d_{jk}$ and $d_{kl}$. We need this notion of cycles to define our detour schemes. \\
According to the form of $\bar N$, we will apply one of the following two Detour schemes:
\subsubsection{Detour Scheme 1 (DS 1)} 
$\bar N$ is in the form of (\ref{eqndet31}) for a certain $\{i,j,k,l\}$. In this case, the Detour will be performed over the 3-node cycle. To simplify the notation, we assume 
\begin{equation}\nonumber
\bar N = d_{ij}+d_{jk}+d_{ki}+d_{li}+d_{lj}+d_{lk} > N 
\end{equation}
Now, we need to reduce the terms corresponding to the degrees of freedom over the cycle in the LHS by subtracting $\lambda$, to guarantee:
\begin{equation}\nonumber
(d_{ij}+d_{jk}+d_{ki})-\lambda+d_{li}+d_{lj}+d_{lk} \leq N
\end{equation}
This operation is equivalent to the transmission of $\lambda$-bits via alternative paths (detours). 
 Considering the reverse cycle, the degrees of freedom over it should be increased as: 
\begin{equation}\nonumber
d_{ji}+d_{ik}+d_{kj}\rightarrow d_{ji}+d_{ik}+d_{kj}+2\lambda
\end{equation}
\subsubsection{Detour Scheme 2 (DS 2)}
$\bar N$ is in the form of (\ref{eqndet41}) for a certain $\{i,j,k,l\}$. In this case, the Detour will be performed through the two 3-node cycles. As in the previous case, we assume that $\max(d_{jl},d_{lj})+\max (d_{ik},d_{ki})$ = $d_{lj}+d_{ik}$
\begin{equation}\nonumber
\bar N=d_{ij}+d_{jk}+d_{kl}+d_{li}+d_{ik}+d_{lj}> N
\end{equation}
First, we should define the 3-nodes cycles in $\bar N$
\begin{equation}\nonumber
\begin{aligned}
d_{kl} \rightarrow d_{lj} \rightarrow d_{jk}
& &
d_{kl} \rightarrow d_{li} \rightarrow d_{ik}
\end{aligned}
\end{equation}
We need to reduce the terms over these cycles, therefore we have to subtract an integer $\alpha$ from them, such that
\begin{equation}\nonumber
d_{ij}+(d_{kl}+d_{li}+d_{ik}+d_{lj}+d_{jk})-\alpha \leq N
\end{equation}
Again, considering the reverse cycles, the degrees of freedom over them should be modified as: 
\begin{equation}\nonumber
d_{lk}+d_{ki}+d_{il}+d_{kj}+d_{jl}\rightarrow d_{lk}+d_{ki}+d_{il}+d_{kj}+d_{jl}+2\alpha
\end{equation}
It remains to show that the resulting network after modification can still achieve DoF using SSA. Once this is proven, we can state the following lemma which is proven in appendix B.
\begin{lemm}
It is possible to achieve any DoF tuple that satisfies the conditions in Theorem 2, with $\bar N>N$, by using one of the two Detour Schemes to convert our network to an equivalent one that achieves the original DoF tuple via alternative paths in the two following cases: 
\begin{itemize}
\item $M_1\neq\min$($N,M_1,M_2+M_3+M_4$), $N \leq M_1+M_2$ and $\bar N$ is not in the form of (\ref{eqndet31}) for $l=1$.
\item $N \leq M_2$.
\end{itemize}
\end{lemm}
\subsection{For the remaining cases}
The SSA and DS can only be used under the conditions stated in Lemmas 3 and 4. We will verify that by the following example.\newline
Consider a network with $(M_1,M_2,M_3,M_4,N)$ = (2,2,2,2,4) and a degrees of freedom tuple $d$ = $(d_{12}$,  $d_{13}$, $d_{14}$, $d_{21}$, $d_{23}$, $d_{24}$, $d_{31}$, $d_{32}$, $d_{34}$, $d_{41}$, $d_{42}$, $d_{43})$ = (1,1,0,0,2,0,0,0,2,0,2,0), this DoF tuple satisfies the conditions in Theorem 2 with $\bar N$ is in the form of (\ref{eqndet31}) for $l=1$ and equal to 6. Thus, we need to detour at least two bits and there is no possible detour can result in an equivalent DoF that satisfies the conditions in Theorem 2, whether the outer bound is achievable for the remaining cases needs more investigation. 
\section{The Outer Bound on the DoF region based on the one sided genie}\label{upperboundsection}
In this section, we explain briefly the development of the outer bound on the DoF region. The authors in \cite{chaaban2013degrees}, proved the converse of the total DoF bound using the traditional cut set bounds \cite{cover2006elements}, and genie-aided bounds. In \cite{tian2013degrees}, an outer bound for the MIMO relay network was developed. Also, we developed an outer bound on the capacity region of the deterministic Y channel and the 4-user relay networks in \cite{zewail2013deterministicitw}, \cite{zewail2013deterministic}, respectively, based on the notion of single sided genie. Due to space limitation, we will not state that here, as it is known that the degrees of freedom between user $i$ and $j$ is defined as 
\begin{equation}\nonumber
d_{ij}=\lim_{P\to \infty} \dfrac{R_{ij}}{0.5 \log P}
\end{equation}
Therefore, the capacity region is directly linked to the degrees of freedom region. Now, we will mention only to the development of the bound on the total degrees of freedom ($D_t$) for the 4-user relay networks. From (\ref{onenode}), by setting $i$= 1 through 4, and noting that since $M_1$ is the largest, then $\min(M_i,N,M_1+M_k+M_l)=\min(M_i,N)$ for $i\geq 2$, we can obtain the following conditions 
\begin{equation}\nonumber
d_{12}+d_{13}+d_{14}\leq \min(M_1,N,M_2+M_3+M_4)
\end{equation}
\begin{equation}\nonumber
d_{21}+d_{23}+d_{24}\leq \min(M_2,N)
\end{equation}
\begin{equation}\nonumber
d_{31}+d_{32}+d_{34}\leq \min(M_3,N)
\end{equation}
\begin{equation}\nonumber
d_{41}+d_{42}+d_{43}\leq \min(M_4,N)
\end{equation}
Adding the above 4 inequalities, we obtain:
\begin{multline}\label{Dt1}
D_t\leq \min(M_1,N,M_2+M_3+M_4)\\+\min(M_2,N)+\min(M_3,N)+\min(M_4,N)
\end{multline}
However, from (\ref{threeUL}) and (\ref{threeDL}), we have
\small
\begin{equation}\label{upper1}
d_{21}+d_{31}+d_{41}+d_{23}+d_{24}+d_{34}\leq \min(N,M_2+M_3+M_4)
\end{equation}
\begin{equation}\label{upper2}%
d_{12}+d_{13}+d_{14}+d_{32}+d_{42}+d_{43}\leq \min(N,M_2+M_3+M_4)
\end{equation}
\normalsize
Again, by summing (\ref{upper1}) and (\ref{upper2}), we get
\begin{equation}\label{Dt2}
D_t\leq 2\min(N,M_2+M_3+M_4)
\end{equation}
By combining (\ref{Dt1}) and (\ref{Dt2}), we get condition (\ref{Dt}).
\section{Numerical Examples}\label{ex}
In this section, we provide numerical examples to illustrate our detour schemes for the Y channel and the 4-user relay networks.
\subsection{An example for the Y channel}
Consider a network with $(M_1,M_2,M_3,N)$ = (3,2,2,3) and $d$ $= (d_{12},d_{13},d_{21},d_{23},d_{31},d_{32})$ = (2,0,0,1,1,0), which satisfies the conditions in Theorem 1, with $\bar{N}=2+1+1=3+1$. Therefore, we will apply the Detour Scheme as follows 
\begin{equation}\nonumber
\begin{aligned}
d_{23} \rightarrow d_{23}-1
& &
d_{21} \rightarrow d_{21}+1
& &
d_{13} \rightarrow d_{13}+1
\end{aligned}
\end{equation}  
Thus, we have a new DoF tuple $d_{n}$=(2,1,1,0,1,0) that satisfies all the conditions stated in Theorem 1 with $\bar{N_{n}}=2+1=3$. Thus, we can apply the SSA to achieve $d_{n}$.
\subsection{Examples for the 4-user relay network}
Consider a network with a degrees of freedom tuple $d$ = $(d_{12}$,  $d_{13}$, $d_{14}$, $d_{21}$, $d_{23}$, $d_{24}$, $d_{31}$, $d_{32}$, $d_{34}$, $d_{41}$, $d_{42}$, $d_{43})$. 
\subsubsection*{Example 1}
Consider a network with $(M_1$, $M_2$, $M_3$, $M_4,N)$ = (6,5,4,3,6) and a degrees of freedom tuple $d$ = (1,1,0,0,1,2,0,0,1,2,0,0), which satisfies the conditions in Theorem 2, with $\bar{N}=2+1+2+1+1+1=6+2$, which is in the form of (\ref{eqndet41}). Therefore, we will apply Detour Scheme 2 as follows 
\begin{equation}\nonumber
\begin{aligned}
d_{24} \rightarrow d_{24}-1
& &
d_{21} \rightarrow d_{21}+1
& &
d_{14} \rightarrow d_{14}+1
\\
d_{34} \rightarrow d_{34}-1
& &
d_{31} \rightarrow d_{31}+1
& &
d_{14} \rightarrow d_{14}+1
\end{aligned}
\end{equation}  
Thus, we have a new DoF tuple $d_{n}$=(1,1,2,1,1,1,1,0,0,2,0,0) that satisfies all the conditions stated in Theorem 2 with $\bar{N_{n}}=6$. Thus, we can apply the SSA to achieve $d_{n}$.
\subsubsection*{Example 2}
Consider a network with $(M_1$, $M_2$, $M_3$, $M_4,N)$ = (6,6,4,3,6), and a degrees of freedom tuple $d$ = (1,1,1,0,2,0,0,0,1,0,1,0), which satisfies the conditions in Theorem 2, with $\bar{N}=1+1+1+2+1+1=6+1$, which is in the form of (\ref{eqndet31}) with $l=1$. Therefore, we will apply Detour Scheme 1 as follows 
\begin{equation}\nonumber
\begin{aligned}
d_{23} \rightarrow d_{23}-1
& &
d_{24} \rightarrow d_{24}+1
& &
d_{43} \rightarrow d_{43}+1
\end{aligned}
\end{equation}  
Thus, we have a new DoF tuple $d_{n}$=(1,1,1,0,1,1,0,0,1,0,1,1) that satisfies all the conditions stated in Theorem 2 with $\bar{N_{n}}=6$. Thus, we can apply the SSA to achieve $d_{n}$.
\section{Conclusion}\label{concl}
In this work, we obtained a larger achievable DoF region for MIMO Y channel by applying a combination of SSA and a detour scheme. The detour scheme employed here is motivated by a similar technique used to achieve the capacity of the deterministic Y channel. These results were extended to 4-user MIMO relay networks. Due to the inherent proximity between the definitions of DoF and deterministic capacity, we envision our results as a first step towards revisiting techniques employed to achieve the deterministic capacity to assist in achieving DoF. More work is needed, however, for a complete characterization of the DoF region for both the Y-channel and the 4 user relay networks.  
\appendices  
\section{Proof of Lemma 2}
Assume that $\bar N$ takes the form in (\ref{ydetour1}), then we have:
\begin{equation}\label{yvioNbar}
\bar{N}=d_{12}+d_{23}+d_{31}=N+\lambda
\end{equation} 
Furthermore, the inequalities in (\ref{yUL2}) and (\ref{yDL2}) can easily be shown to imply the following:
\begin{equation}\label{extra1}
d_{12}+d_{13}+d_{23}\leq N
\end{equation} 
\begin{equation}\label{extra2}
d_{21}+d_{23}+d_{31} \leq N
\end{equation}
\begin{equation}\label{extra3}
d_{12}+d_{32}+d_{31} \leq N
\end{equation}  
Using (\ref{yvioNbar}) to substitute in the LHS of (\ref{extra1})-(\ref{extra3}) we obtain:
\begin{equation}\label{yinsights}
\begin{aligned}
d_{12} \geq d_{21}+\lambda
& & d_{23} \geq d_{32}+\lambda
& & d_{31} \geq d_{13}+\lambda
\end{aligned}
\end{equation}
Now, will apply the DS as follows
\begin{equation}\nonumber
\begin{aligned}
d_{23} \rightarrow d_{23}-\lambda
& & d_{21} \rightarrow d_{21}+\lambda
& & d_{13} \rightarrow d_{13}+\lambda
\end{aligned}
\end{equation}
From (\ref{yinsights}) and under the condition $M_1\neq\min$($N$,$M_1$,$M_2+M_3)$, we can verify that the conditions in Theorem 1 are satisfied for the modified DoF tuple and  $\bar N =N$.\\
Note that we do not need to check any condition that is restricted by $N$, because we guarantee that it will be satisfied as $\bar N$ is equal to the largest possible value over all LHS's.
\section{Proof of Lemma 4}
\subsection{Detour Scheme 1}
If we assume $\bar N$ in the following formula:  
\begin{equation}\label{M1NbarDS1}
\bar N= d_{13}+d_{34}+d_{41}+d_{12}+d_{32}+d_{42}=N+\lambda
\end{equation}
However, from (\ref{threeDL}) and (\ref{threeUL}) in Theorem 2, we can get  
\begin{equation}\nonumber
d_{31}+d_{34}+d_{32}+d_{41}+d_{42}+d_{12}\leq N
\end{equation}
\begin{equation}\nonumber
d_{41}+d_{42}+d_{43}+d_{12}+d_{13}+d_{32}\leq N
\end{equation}
\begin{equation}\nonumber
d_{13}+d_{12}+d_{14}+d_{32}+d_{34}+d_{42}\leq N
\end{equation}
By comparing the above conditions with (\ref{M1NbarDS1}), we get
\begin{equation}\label{eqn: R1icondUB}
\begin{aligned}
d_{13} \geq d_{31}+\lambda
& & d_{34} \geq d_{43}+\lambda
& & d_{41} \geq d_{14}+\lambda
\end{aligned}
\end{equation} 
We will apply the detour scheme, in which we detour $\lambda$ bits from node 4 to node 1 via node 3, thus the DoF over this cycle will be modified as follows:
\begin{equation}\nonumber
\begin{aligned}
d_{34} \rightarrow d_{34}-\lambda
& & d_{31} \rightarrow d_{31}+\lambda
& & d_{14} \rightarrow d_{14}+\lambda
\end{aligned}
\end{equation}
From (\ref{M1NbarDS1}) and (\ref{eqn: R1icondUB}), for the modified DoF tuple, we have 
\begin{multline}\label{M1NbarDS122}
\bar N= \max(d_{13},d_{31}+\lambda)+\max(d_{41},d_{14}+\lambda)\\+\max(d_{34}-\lambda,d_{43})
+d_{12}+d_{32}+d_{42}=N
\end{multline}
Then, we need to check the conditions that contain $d_{31}$ or $d_{14}$ in Theorem 2 for the modified DoF tuple. It is clear that from (\ref{M1NbarDS122}), all conditions restricted by $N$ are now satisfied, since its LHS has the largest possible value among all conditions. From (\ref{eqn: R1icondUB}) and under the conditions $M_1\neq\min$($N,M_1,M_2+M_3+M_4$) and $N \leq M_1+M_2$, we can verify that all conditions in Theorem 2 are now satisfied, therefore we can apply the SSA.  
\subsection{Detour Scheme 2}
If we assume $\bar N$ in the following formula:
\begin{equation}\label{M1DS2}
\bar N=d_{41}+d_{12}+d_{24}+d_{13}+d_{34}+d_{23}=N+\lambda
\end{equation} 
which contains two 3-node cycles as follows:
\begin{equation}\nonumber
\begin{aligned}
d_{41}\rightarrow d_{13}\rightarrow d_{34}
& & &
d_{41}\rightarrow d_{12}\rightarrow d_{24}
\end{aligned}
\end{equation}
However, from (\ref{threeUL}) in Theorem 2, we can get
\begin{equation}\nonumber
d_{14}+d_{12}+d_{13}+d_{23}+d_{24}+d_{34}\leq N
\end{equation} 
\begin{equation}\nonumber
d_{21}+d_{23}+d_{24}+d_{41}+d_{43}+d_{13}\leq N
\end{equation} 
By comparing these conditions with (\ref{M1DS2}), we get
\begin{equation}\label{eqn: R4condUB}
\begin{aligned}
d_{41} &\geq d_{14}+\lambda
\\  d_{34}+d_{12} &\geq d_{43}+d_{21}+\lambda
\end{aligned} 
\end{equation}
Also, from (\ref{M1DS2}), we can let 
\begin{equation}\label{eq1}
d_{41}+d_{12}+d_{24}+d_{13}+d_{43}+d_{23}=N+\beta
\end{equation}
\begin{equation}\label{eq2}
d_{41}+d_{13}+d_{34}+d_{21}+d_{24}+d_{23}=N+\gamma
\end{equation}
By comparing these conditions with (\ref{M1DS2}), we get
\begin{equation}\label{eqn: R4condSOS1}
\begin{aligned}
d_{34}=d_{43}+\lambda-\beta
& & d_{12}=d_{21}+\lambda-\gamma
\end{aligned}
\end{equation} 
Again, from the conditions in Theorem 2, we get 
\begin{equation}\label{eq11}
d_{41}+d_{12}+d_{42}+d_{13}+d_{43}+d_{23}\leq N
\end{equation}
\begin{equation}\label{eq22}
d_{41}+d_{31}+d_{34}+d_{21}+d_{24}+d_{23}\leq N
\end{equation}
And by comparing (\ref{eq1}), (\ref{eq2}), (\ref{eq11}) and (\ref{eq22}), we get 
\begin{equation}\label{eqn: R4condSOS2}
\begin{aligned}
d_{24}\geq d_{42}+\beta
& &  d_{13}\geq d_{31}+\gamma
\end{aligned}
\end{equation} 
Now, we will apply DS as follows  
\begin{equation}\nonumber
\begin{aligned}
d_{24} \rightarrow d_{24}-\beta
& &
d_{21} \rightarrow d_{21}+\beta
& &
d_{14} \rightarrow d_{14}+\beta
\\
d_{34} \rightarrow d_{34}-\gamma
& &
d_{31} \rightarrow d_{31}+\gamma
& &
d_{14} \rightarrow d_{14}+\gamma
\end{aligned}
\end{equation}
and from (\ref{eqn: R4condUB}) and (\ref{eqn: R4condSOS1}), we guarantee that $\lambda = \beta+\gamma$. \newline
From (\ref{M1DS2}), (\ref{eqn: R4condSOS1}) and (\ref{eqn: R4condSOS2}), for the modified DoF tuple, we have $\bar N = N$. Again, we need to check the conditions in Theorem 2 for the modified DoF tuple, and from (\ref{eqn: R4condSOS1}) and (\ref{eqn: R4condSOS2}) and under the conditions $M_1\neq\min$($N,M_1,M_2+M_3+M_4$) and $N \leq M_1+M_2$, we can verify that those conditions are satisfied. 
\subsection{$N \leq M_2$}
Since, $M_1 \geq M_2$ and $N \leq M_2$, all above Detours will work in this case, therefore we need only to prove the validity of the Detour when $\bar{N}$ is in the form (\ref{eqndet31}) for $l=1$. The proof follows the same steps as the proof of Detour Scheme 1 that we explained before.    
\bibliographystyle{IEEEtran}
\bibliography{IEEEabrv,DiversityLib}

\end{document}